\documentclass[11pt]{article}
\pdfoutput=1
\usepackage{amsmath,amsthm}
\usepackage[noadjust]{cite}
\usepackage{graphicx}
\usepackage{listings}
\usepackage{multirow}
\usepackage{color}
\usepackage{booktabs}
\usepackage{url}
\usepackage[caption=false,font=footnotesize]{subfig}
\usepackage{fullpage}
\usepackage{tabularx}

\title{Parallel structurally-symmetric sparse matrix-vector products\\
on multi-core processors~\thanks{This work was partially supported by CNPq, Brazil.}}

\author{%
Vicente H.~F.~Batista\thanks{Programa de Engenharia Civil, Universidade Federal do Rio de
Janeiro, COPPE, Caixa Postal 68506, Rio de Janeiro, RJ, 21945-970, Brazil.
Email: {\tt \{helano,george,fernando\}@coc.ufrj.br}.} \and
George O.~Ainsworth Jr.\footnotemark[2] \and
Fernando L.~B.~Ribeiro\footnotemark[2]}

\begin{document}

\lstset{language=Fortran,basicstyle=\footnotesize\tt,numbers=left,numberstyle=\tt\scriptsize,keywordstyle=\tt\bf}

\maketitle

\begin{abstract}

We consider the problem of developing an efficient multi-threaded
implementation of the matrix-vector multiplication algorithm for sparse
matrices with structural symmetry.  Matrices are stored using the
\textit{compressed sparse row-column} format (CSRC), designed for profiting
from the symmetric non-zero pattern observed in global finite element matrices.
Unlike classical compressed storage formats, performing the sparse
matrix-vector product using the CSRC requires thread-safe access to the
destination vector.  To avoid race conditions, we have implemented two
partitioning strategies.  In the first one, each thread allocates an array for
storing its contributions, which are later combined in an accumulation step.
We analyze how to perform this accumulation in four different ways.
The second strategy employs a coloring
algorithm for grouping rows that can be concurrently processed by threads. Our
results indicate that, although incurring an increase in the working set size,
the former approach leads to the best performance improvements for
most matrices.

\bigskip


\end{abstract}

\section{Introduction}

It is not feasible anymore to expect performance gains for sequential codes by means of continuously
increasing processor clock speeds. Nowadays, processor vendors have been
concentrated on developing systems that group two or more processors onto a
single socket, sharing or not the same memory resources.  This technology, called
\textit{multi-core}, has been successfully employed to different application
domains ranging from computer graphics to scientific computing, and in these
times it is commonly seen on high performance clusters, desktop
computers, notebooks, and even mobile devices.  The spread of such architecture
has consequently stimulated an increasing number of researches on parallel
algorithms.

To obtain efficient implementations of parallel algorithms, one must consider
the underlying architecture on which the program is supposed to be run.  In
fact, even processors belonging to the multi-core family may present different
hardware layouts, which can make an implementation to perform poorly on one
platform, while running fast on another.  As an example of such issue,
multi-core processors may have different memory subsystems for each core, therefore
forcing programmers to take care of thread and memory affinity.

The finite element method is usually the first choice for numerically solving
integral and partial differential equations.  Matrices arising from finite
element discretizations are usually sparse, i.e., most of its entries are
zeros.  Effectively storing sparse matrices requires the use of compressed data
structures.  Commonly employed approaches are the \textit{element-based},
\textit{edge-based} and \textit{compressed} data
structures.  Since the later provides the best compromise between space
complexity and performance \cite{RC05a}, it was chosen as the primary data
structure of our implementation.

The \textit{compressed sparse row} (CSR) data structure stores contiguously in memory non-zero entries belonging
to the same row of a matrix.  While in a dense representation any
element can be randomly accessed through the use of its row and
column indices, the CSR explicitly stores in memory the combinatorial
information for every non-zero entry.  Given an $n \times n$ matrix $A$ with $nnz$
non-zero coefficients, the standard version of the CSR \cite{Saa95a} consists
of three arrays: two integer arrays $ia(n+1)$ and $ja(nnz)$ for storing
combinatorial data, and one floating-point array $a(nnz)$ containing the
non-zero coefficients. The value $ia(i)$ points to the first element of row $i$
in the $a$ array, i.e., row $i$ is defined as the subset of $a$ starting and
ending at $ia(i)$ and $ia(i+1)-1$, respectively.  The column index of each
non-zero entry is stored in $ja$.  There is also a transpose version,
called \textit{compressed sparse column} (CSC) format.

This representation supports matrices of arbitrary shapes and symmetry
properties.  In the context of the finite element method, however, the
generality provided by the CSR is underused as most matrices are structurally symmetric.
In this case, it would be sufficient to store, roughly, half
of the matrix connectivity. The \textit{compressed sparse row-column} (CSRC)
format was designed to take benefit from this fact \cite{RF07a}.  Basically, it
stores the column indices for only half of the off-diagonal entries.  As the
working set size has a great impact on the performance of CSR-like data
structures, the running time of algorithms such as the matrix-vector product is expected
to be improved when using the CSRC.  Also, solvers based on oblique projection methods can efficiently
access the transpose matrix, since it is implicitly defined.

The performance of finite element codes using iterative
solvers is dominated by the computations associated with the matrix-vector
multiplication algorithm. In this algorithm, we are given an $n \times n$
sparse matrix $A$ containing $nnz$ non-zeros, and a dense $n$-vector $x$,
called the \textit{source} vector.  The output is an $n$-vector $y$, termed the
\textit{destination} vector, which stores the result of the $Ax$ operation.
Performing this operation using the CSR format is trivial, but it was observed
that the maximum performance in Mflop/s sustained by a na\"ive implementation
can reach only a small part of the machine peak performance \cite{GKKS99a}.  As
a means of transcending this limit, several optimization techniques have been
proposed, such as reordering \cite{Tol97a,PH99a,WS97a,TJ92a}, data compression \cite{MGMM05a,WL06a},
blocking \cite{TJ92a,IYV04a,Tol97a,VM05a,PH99a,AGZ92a,NVDY07a}, vectorization \cite{AFM05a,BHZ93a}, loop unrolling
\cite{WS97a} and jamming \cite{MG04a}, and software prefetching \cite{Tol97a}.
Lately, the dissemination of multi-core computers have promoted
multi-threading as an important tuning
technique, which can be further combined with purely sequential methods.

\subsection{Related work}

Parallel sparse matrix-vector multiplication using CSR-like data structures on
multi-processed machines has been the focus of a number of researchers since
the 1990s.  Early attempts to date include the paper by {\c C}ataly\"urek and Aykanat \cite{CA96a}, on
hypergraph models applied to the matrix partitioning problem, Im and Yelick \cite{IY99a},
who analysed the effect of register/cache blocking and reordering, and
Geus and R{\"{o}}llin \cite{GR01a}, considering prefetching, register blocking and reordering for
symmetric matrices.  Kotakemori et al.~\cite{KHKNSN05a} also examined several storage formats on
a ccNUMA machine, which required the ability of dealing with page allocation
mechanisms.

Regarding modern multi-core platforms, the work of Goumas et al.~\cite{GKAKK08a} contains a
thorough analysis of a number of factors that may degrade the performance of
both sequential and multi-thread implementations.  Performance tests were
carried out on three different platforms, including SMP, SMT and ccNUMA
systems.
Two partitioning schemes were implemented, one guided by the number of rows
and the other by the number of non-zeros per thread.  It was
observed that the later approach contributes to a better load balancing,
thus improving significantly the running time.
For large matrices, they obtained average speedups of 1.96 and 2.13 using 2 and
4 threads, respectively, on an Intel Core 2 Xeon.
In this platform, their code reached about 1612 Mflop/s for 2 threads,
and 2967 Mflop/s when spawning 4 threads.
This performance changes
considerably when considering matrices whose working set sizes are far from fitting in cache.
In particular, it drops to around 815 Mflop/s and 849 Mflop/s, corresponding to the 2-
and 4-threaded cases.

Memory contention is viewed as the major bottleneck of implementations of the
sparse matrix-vector product. This problem was tackled by Kourtis et al.~\cite{KGK08a} via
compression techniques, reducing both the matrix connectivity and
floating-point numbers to be stored.  Although leading to good scalability,
they obtained at most a 2-fold speedup on 8 threads, for matrices out of cache.
The experiments were conducted on two Intel Clovertown with 4MB of L2
cache each.  In the same direction, Belgin et al.~\cite{BBR09a} proposed a
pattern-based blocking scheme for reducing the index overhead.
Accompanied by software prefetching and vectorization techniques, they attained
an average sequential speedup of 1.4.  Their multi-thread implementation required the
synchronization of the accesses to the $y$ vector. In brief, each thread
maintains a private vector for storing partial values, which are summed up in a
reduction step into the global destination vector.  They observed average
speedups around 1.04, 1.11 and 2.3 when spawning 2, 4, and 8 threads,
respectively.  These results were obtained on a 2-socket Intel Harpertown 5400
with 8GB of RAM and 12MB L2 cache per socket.

Different row-wise partitioning methods were considered by Liu et al.~\cite{LZSQ09a}.
Besides evenly splitting non-zeros among threads, they evaluated
the effect of the automatic scheduling mechanisms provided by OpenMP, namely,
the \textit{static}, \textit{dynamic} and \textit{guided} schedules.  Once
more, the non-zero strategy was the best choice.  They also
parallelized the block CSR format.  Experiments were run on four AMD Opteron 870
dual-core processors, with 16GB of RAM and $2 \times 1$MB L2 caches.  Both
CSR and block CSR schemes resulted in poor scalability for large matrices,
for which the maximum speedup was approximately 2, using 8 threads.

Williams et al.~\cite{WOVSYD09a} evaluated the sparse matrix-vector kernel using the CSR
format on several up-to-date chip multiprocessor systems, such as the
heterogeneous STI Cell.  They examined the effect of various optimization techniques
on the performance of a multi-thread CSR, including software
pipelining, branch elimination, SIMDization, explicit prefetching, 16-bit
indices, and register, cache and translation lookaside buffer (TLB) blocking.
A row-wise approach was employed for
thread scheduling.  As regarding finite element matrices and in comparison to OSKI~\cite{VDY05a},
speedups for the fully tuned parallel code ranged
from 1.8 to 5.5 using 8 threads on an Intel Xeon E5345.

\begin{figure}[!t]
\centering
\includegraphics{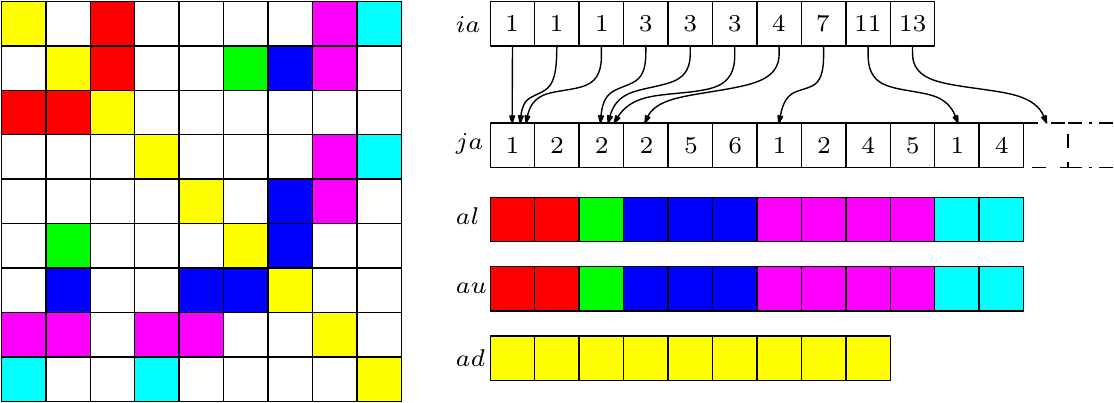}
\caption{The layout of CSRC for an arbitrary 9$\times$9 non-symmetric matrix.}
\label{fig:csrc_scheme}
\end{figure}

More recently, Bulu{\c{c}} et al.~\cite{BFFGL09a} have presented a block structure that allows
efficient computation of both $Ax$ and $A^{\mathsf{T}}x$ in parallel.  It can
be roughly seen as a dense collection of sparse blocks, rather than a sparse
collection of dense blocks, as in the standard block CSR format.  In
sequential experiments carried out on an ccNUMA machine featuring AMD Opteron 8214
processors, there were no improvements over the standard CSR.
In fact, their data structure was always slower for band matrices.  Concerning
its parallelization, however, it was proved that it yields a
parallelism of $\Theta(nnz/\sqrt{n}\log n)$.
In practice, it scaled up to 4 threads on an Intel Xeon X5460, and
presented linear speedups on an AMD Opteron 8214 and an Intel Core i7 920.  On
the later, where the best results were attained, it reached speedups of 1.86,
2.97 and 3.71 using 2, 4 and 8 threads, respectively. However,
it does not seem to straightly allow the simultaneous
computation of $y_i \leftarrow y_i + a_{ij} x_j$ and $y_j \leftarrow y_j +
a_{ij} x_i$ in a single loop, as CSRC does.

\subsection{Overview}

The remainder of this paper is organized as follows.  Section \ref{sec:csrc}
contains a precise definition of the CSRC format accompanied with a description
of the matrix-vector multiplication algorithm using such structure.  Its
parallelization is described in Section \ref{sec:parallel-csrc}, where we
present two strategies for avoiding conflicts during write accesses to
the destination vector.  Our results are shown in Section \ref{sec:results},
supplemented with some worthy remarks.  We finally draw some conclusions in
Section \ref{sec:conclusion}.

\section{The CSRC storage format}
\label{sec:csrc}

The \textit{compressed sparse row-column} (CSRC) format is a specialization of
CSR for structurally symmetric matrices arising in finite element modelling
\cite{RF07a}, which is the target domain application of this work.  Given an
arbitrary $n \times n$ global matrix $A = (a_{ij})$, with $nnz$
non-zeros, the CSRC decomposes $A$ into the sum $A_D + A_L + A_U$, where $A_D$,
$A_L$, and $A_U$ correspond to the diagonal, lower and upper parts of $A$,
respectively.  The sub-matrix $A_L$ (resp.~$A_U$) is stored in a row-wise
(resp.~column-wise) manner.

In practice, the CSRC splits the off-diagonal coefficients into two
floating-point arrays, namely, $al(k)$ and $au(k)$, $k = \frac{1}{2}(nnz - n)$,
where the lower and upper entries of $A$ are stored.  In other words, if $j <
i$, then $a_{ij}$ is stored in $al$, and $au$ contains its transpose $a_{ji}$.
The diagonal elements are stored in an array $ad(n)$.  Other two integer
arrays, $ia(n+1)$ and $ja(k)$, are also maintained. These arrays can be defined
in terms of either the upper or lower coefficients.  The $ia$ array
contains pointers to the beginning of each row (resp.~column) in $al$
(resp.~$au$), and $ja$ contains column (resp.~row) indices for those non-zero
coefficients belonging to $A_L$ (resp.~$A_U$).  Another
interpretation is that $A_L$ is represented using CSR, while $A_U$ is stored using
CSC.  We illustrate the CSRC data structure for an arbitrary 9$\times$9 non-symmetric matrix
consisting of 33 non-zeros in Figure \ref{fig:csrc_scheme}.

Notice that the CSRC could be viewed as the sparse skyline (SSK) format
restricted to structurally symmetric matrices \cite{Saa95a,GR01a}.
However, as shown in Section \ref{sec:rectextension}, we made it capable of
representing rectangular matrices after minor modifications.  Furthermore, to
our knowledge, this is the first evaluation of such structure on modern
multi-processed machines.

\subsection{Extension to rectangular matrices}
\label{sec:rectextension}

The way the CSRC is defined would disallow us handling matrices with different
aspect ratios other than square.  In the overlapping strategy
implemented in any distributed-memory finite element code using a 
subdomain-by-subdomain approach
\cite{RF07a,ARM09a}, it is normal the
occurrence of rectangular matrices with a remarkable property.
An $n \times m$ matrix $A$,
with $m > n$, can always be written as the sum $A_S + A_R$, where
$A_S$ and $A_R$ are of order $n \times n$ and $n \times k$, respectively, with
$k = m - n$.  In addition, the $A_S$ matrix has symmetric non-zero pattern, and
it is occasionally numerically symmetric.  Therefore, it can be represented by the
CSRC definition given before, while $A_R$ can be stored using an auxiliary CSR
data structure.

\begin{figure}[!t]
\centering
\subfloat[]{\label{fig:matvec_csrc}%
\begin{lstlisting}[boxpos=b]
do i = 1, n
   xi = x(i)
   t  = ad(i)*xi
   do k = ia(i), ia(i+1)-1
      jak = ja(k)
      t   = t + al(k)*x(jak)
      y(jak) = y(jak) + au(k)*xi
   enddo
   y(i) = t
enddo
\end{lstlisting}}
\hfil
\subfloat[]{\label{fig:matvec_csrcr}%
\begin{lstlisting}[boxpos=b]
do i = 1, n
   xi = x(i)
   t  = ad(i)*xi
   do k = ia(i), ia(i+1)-1
      jak = ja(k)
      t   = t + al(k)*x(jak)
      y(jak) = y(jak) + au(k)*xi
   enddo
   do k = iar(i), iar(i+1)-1
      jak = jar(k)
      t   = t + ar(k)*x(jak)
   enddo
   y(i) = t
enddo
\end{lstlisting}}
\caption{Code snippets for the non-symmetric matrix-vector multiplication
algorithm using CSRC for (a) square and
(b) rectangular matrices.}
\label{fig:SpMV}
\end{figure}

\subsection{Sequential matrix-vector product}

The sequential version of the CSRC matrix-vector multiplication algorithm has
the same loop structure as for CSR.  The input matrix $A$ is
traversed by rows, and row $i$ is processed from the left to the right up to
its diagonal element $a_{ii}$.  Because we assume $A$ is structurally
symmetric, its upper part can be simultaneously traversed.  That is, we are
allowed to compute both $y_i \leftarrow y_i + a_{ij} x_j$ and $y_j \leftarrow
y_j + a_{ji} x_i$, in the $i$-th loop.  If $A$ is also numerically symmetric, we can further
eliminate one load instruction when retrieving its upper entries.  For
rectangular matrices, there is another inner loop to process the coefficients
stored in the auxiliary CSR.  Figure \ref{fig:SpMV} contains Fortran implementations of
the sparse matrix-vector product using CSRC for square and rectangular matrices.

\section{Parallel implementation}
\label{sec:parallel-csrc}

To parallelize the sparse matrix-vector product using the CSRC, one can basically spawn threads at either the inner or the outer loop.
This means adding a \texttt{parallel do} directive just above line 1 or 4 of Figure \subref{fig:matvec_csrc} (and 9, for Figure \subref{fig:matvec_csrcr}).
As the amount of computations per row is usually low, the overhead due to the inner parallelization would counteract any parallelism.
On the other hand, recall that the CSRC matrix-vector product has the property that
the lower and upper parts of the input matrix are simultaneously traversed.
Thus spawning threads at line 1 requires the synchronization of writings into the destination vector.
That is, there exists a race condition on the access of the vector $y$.
If two threads work on different rows,
for example, rows $i$ and $j$, $j > i$, it is not unlikely that both threads
require writing permission to modify $y(k)$, $k \leq i$.

In short, our data structure is required to support concurrent reading and
writing on the vector $y$.  These operations need to be thread-safe, but at
the same time very efficient, given the fine granularity of the operations.
Common strategies to circumvent this problem would employ atomic primitives,
locks, or the emerging transactional memory model.  However, the overheads incurred by
these approaches are rather costly, compared to the total cost of accessing
$y$.  A more promising solution would be to determine subsets of rows
that can be handled by distinct threads in parallel.  In this paper, we have
considered two of such solutions, here termed \textit{local buffers} and
\textit{colorful} methods.

Our algorithms were analyzed using the concepts of \textit{work} and
\textit{span} \cite[Ch.~27]{CLRS09a}.  The \textit{work} $T_{1}$ of an
algorithm is the total cost of running it on exactly one processor, and the
\textit{span} $T_{\infty}$ is equal to its cost when running on an infinite
number of processors.  The \textit{parallelism} of a given algorithm is then
defined as the ratio $T_{1}/T_{\infty}$.  So, the greater the parallelism of
an algorithm, the better the theoretical guarantees on its performance.  The
work of the matrix-vector multiply using the CSRC is clearly $\Theta(nnz)$.  To
calculate its span, we need to consider our partitioning strategies
separately.

\subsection{Local buffers method}

\begin{figure}[t]
\centering
\subfloat[]{\includegraphics{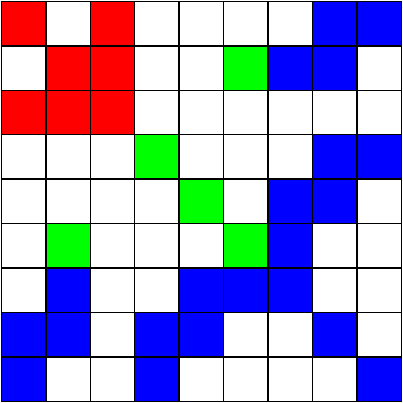}
\label{fig:partitioning-simple}}
\hfil
\subfloat[]{\includegraphics{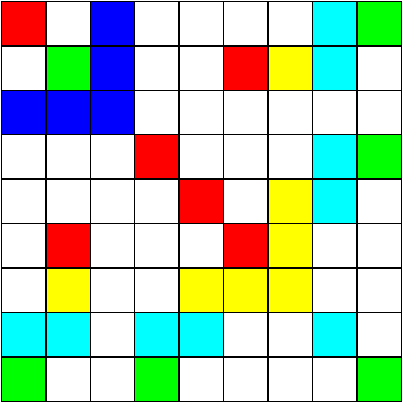}
\label{fig:partitioning-colorful}}
\hfil
\subfloat[]{\includegraphics{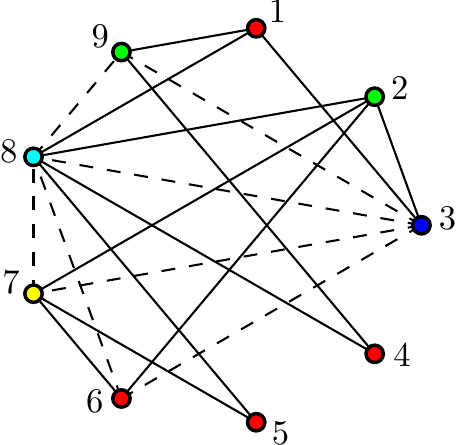}
\label{fig:conflicts}}
\caption{Illustration of the (a) local buffers and the (b) colorful
partitioning methods for 3 threads on a $9 \times 9$ matrix along with its (c) conflict graph.}
\label{fig:partitioning}
\end{figure}

One way to avoid conflicts at the $y$ vector is to assign different
destination vectors to each thread.  That is, thread $t_i$ would compute its
part of the solution, store it in a local buffer $y_i$, and then accumulate
this partial solution into the $y$ vector.  This method, here called
\textit{local buffers method}, is illustrated in
Figure \subref{fig:partitioning-simple}, which shows the distribution of rows for an
arbitrary non-symmetric $9 \times 9$ matrix.  In the example, the matrix is
split into three regions to be assigned to three different threads.  The number of non-zeros per thread is 7, 5 and
21.

The main drawback of this method is the introduction of two additional steps: initialization and accumulation.
The accumulation step is performed to compute the final destination vector resultant from merging partial values stored in local buffers.
Threads must initialize their own buffers, because of this accumulation, otherwise they would store wrong data.
For convenience, we define the \textit{effective range} of a thread as the set of rows in $y$ that it indeed needs to modify.
We consider four ways of implementing both steps:

\begin{enumerate}

\item \textit{All-in-one}: threads initialize and accumulate in parallel the buffers of the whole team.

\item \textit{Per buffer}: for each buffer, threads initialize and accumulate in parallel.

\item \textit{Effective}: threads initialize and accumulate in parallel over the corresponding effective ranges.

\item \textit{Interval}: threads initialize and accumulate in parallel over intervals of $y$ defined by the intersection of their effective ranges.

\end{enumerate}

The spans of the \textit{all-in-one} and \textit{per buffer} methods are $\Theta(p + \log n)$ and $\Theta(p\log n)$, respectively.
If the number of threads is $\Theta(n)$, then their respective parallelism are $O(nnz / n)$ and $O(nnz/n\log n)$.
The platforms considered herein,
however, feature at most four processors. Our experiments will show that
these methods can still provide reasonable scalability for such systems.
In this case, their parallelism would be better approximated by $O(nnz / \log n)$, as for CSR.

In fact, the problem with the first two methods is that they treat all buffers as dense vectors, which is rarely true in practice as we are dealing with sparse matrices.
The \textit{effective} and \textit{interval} methods try to mitigate this issue by performing computations only on effective ranges.
For narrow band matrices, which is usually the case of finite element matrices, we can assume the effective range is $\Theta(n/p)$.
Hence the span of both methods is $\Theta(p \log (n/p))$.

Since the work per thread strongly depends on the number of non-zeros per row,
a partitioning technique based just on the number of rows may result in load
imbalance.  A more efficient way is to consider the number of non-zeros per
thread, because the amount of floating point operations becomes balanced.  The
results presented herein were obtained using such a non-zero guided implementation,
in which the deviation from the average number of non-zeros per row
is minimized.

\subsection{Colorful method}

The \textit{colorful method} partitions a matrix into sets of pairwise
conflict-free rows.
Here we distinguish between two kinds of conflicts.
If a thread owns row $i$ and a second thread owning row $j$, $j > i$, requires
to modify $y(k)$, $k < i$, it is called an \textit{indirect} conflict.
If $k = i$, we call such conflict \textit{direct}.
The \textit{conflict graph} of a matrix $A$ is the graph $G[A] = (V,E)$,
where each vertex $v \in V$ corresponds to a row in $A$, and the edges in $E$
represent conflicts between vertices.
Figure \subref{fig:conflicts} shows the conflict graph for the matrix in
Figure \ref{fig:csrc_scheme}.
Direct and indirect conflicts are indicated by solid and dashed lines,
respectively.  In the graph, there are 12 direct and 7 indirect conflicts.

The direct
conflicts of row $i$ are exactly the rows corresponding to the column indices of
the non-zero entries at that row, i.e., the indices $ja(k)$, $k \in [ia(i), ia(i+1))$.  They can be computed in a
single loop through the CSRC structure.  The computation of indirect conflicts
is more demanding.  In our implementation, these are determined with the aid of
the induced subgraph $G'[A]$ spanned by the edges in $G[A]$ associated with
direct conflicts.  Given two vertices $u, v \in V$, if the intersection of
their neighborhood in $G'[A]$ is non-empty, then they are indirectly in
conflict.

We color the graph $G[A]$ by applying a standard sequential
coloring algorithm \cite{CM83a}.
The color classes correspond to
conflict-free blocks where the matrix-vector product can be safely carried
out in parallel.  Observe that coloring rectangular matrices
is the same as coloring only its square part, since
the rectangular part is accessed by rows.  The layout of a 5-colored
matrix is depicted in Figure \subref{fig:partitioning-colorful}.

Let $k$
denote the number of colors used by the coloring algorithm.
Suppose that the color classes are evenly sized, and
that the loop over the rows is implemented as a divide-and-conquer recursion.
Under such hypothesis, the span of the colorful method can be approximated by $\Theta(k\log(n/k))$.
Thus, the colorful matrix-vector product has a parallelism of $O(nnz / k\log(n/k) )$.
Although $k < p$ would lead to a better scalability when compared to the local buffers strategy,
the possibility of exploiting systems based on cache hierarchies decreases,
which affects considerably the code performance.
Furthermore, the number of processors used in our experiments was always smaller than the number of colors.

\section{Experimental results}
\label{sec:results}

Our implementation was evaluated on two Intel processors, including an Intel
Core~2 Duo E8200 (codenamed \textit{Wolfdale}) and an Intel i7 940 (codenamed
\textit{Bloomfield}).  The Wolfdale processor runs at 2.66GHz, with L2 cache of
6MB and 8GB of RAM, and the Bloomfield one runs at 2.93GHz with $4\times$256KB
L2 caches, 8MB of L3 cache and 8GB of RAM. Our interest on Intel Core~2 Duo
machines lies on the fact that our finite element simulations are carried out
on a dedicated 32-nodes cluster of such processors.

The code was parallelized using OpenMP directives, and compiled
with Intel Fortran compiler (\texttt{ifort}) version 11.1 with level 3
optimizations (\texttt{-O3} flag) enabled.  Machine counters were accessed
through the PAPI 3.7.1 library API \cite{BDGHM00a}.  The measurements of
speedups and Mflop/s were carried out with PAPI instrumentation disabled.

The tests were performed on a data set comprised of 60 matrices, from which 32 are
numerically symmetric.  There is one non-symmetric dense matrix of order 1K, 50
matrices selected from the University of Florida sparse matrix collection
\cite{Dav97a}, and 3 groups of 3 matrices each, called angical, tracer, and
cube2m, of our own devise.  Inside these groups, matrices correspond to
one global finite element matrix output by our sequential finite element code,
and two global matrices for both of the adopted domain partitioning schemes,
overlapping (suffix ``\_o32'') and non-overlapping (suffix ``\_n32''),
where 32 stands for the number of sub-domains.  Our benchmark computes the sparse
matrix-vector product a thousand times for each matrix in Table
\ref{tab:matrices-details}, which is a reasonable value for iterative solvers
like the preconditioned conjugate gradient method and the generalized
minimum residual method. All results correspond to
median values over three of such runs.

\begin{table}[!t]
\caption{Details of the matrices used in our experiments.}
\label{tab:matrices-details}
\centering
{\scriptsize
\begin{tabularx}{0.485\textwidth}{@{}ll@{\ \ }rrrr@{}}
\toprule
Matrix & Sym. & $n$ & $nnz$ & $nnz/n$ & $ws$ (KB)\\
\midrule
thermal & no & 3456 & 66528 & 19 & 710 \\
ex37 & no & 3565 & 67591 & 18 & 722 \\
flowmeter5 & no & 9669 & 67391 & 6 & 828 \\
piston & no & 2025 & 100015 & 49 & 1012 \\
SiNa & yes & 5743 & 102265 & 17 & 1288 \\
benzene & yes & 8219 & 125444 & 15 & 1598 \\
cage10 & no & 11397 & 150645 & 13 & 1671 \\
spmsrtls & yes & 29995 & 129971 & 4 & 1991 \\
torsion1 & yes & 40000 & 118804 & 2 & 2017 \\
minsurfo & yes & 40806 & 122214 & 2 & 2069 \\
wang4 & no & 26068 & 177196 & 6 & 2188 \\
chem\_master1 & no & 40401 & 201201 & 4 & 2675 \\
dixmaanl & yes & 60000 & 179999 & 2 & 3046 \\
chipcool1 & no & 20082 & 281150 & 14 & 3098 \\
t3dl & yes & 20360 & 265113 & 13 & 3424 \\
poisson3Da & no & 13514 & 352762 & 26 & 3682 \\
k3plates & no & 11107 & 378927 & 34 & 3895 \\
gridgena & yes & 48962 & 280523 & 5 & 4052 \\
cbuckle & yes & 13681 & 345098 & 25 & 4257 \\
bcircuit & no & 68902 & 375558 & 5 & 4878 \\
angical\_n32 & yes & 20115 & 391473 & 19 & 4901 \\
angical\_o32 & no & 18696 & 732186 & 39 & 4957 \\
tracer\_n32 & yes & 33993 & 443612 & 13 & 5729 \\
tracer\_o32 & no & 31484 & 828360 & 26 & 5889 \\
crystk02 & yes & 13965 & 491274 & 35 & 5975 \\
olafu & yes & 16146 & 515651 & 31 & 6295 \\
gyro & yes & 17361 & 519260 & 29 & 6356 \\
dawson5 & yes & 51537 & 531157 & 10 & 7029 \\
ASIC\_100ks & no & 99190 & 578890 & 5 & 7396 \\
bcsstk35 & yes & 30237 & 740200 & 24 & 9146 \\
\bottomrule
\end{tabularx}
\hfill
\begin{tabularx}{0.485\textwidth}{@{}l@{\ \ }l@{\ \ }r@{\ \ }r@{\ \ }r@{\ \ }r@{}}
\toprule
Matrix & Sym. & $n$ & $nnz$ & $nnz/n$ & $ws$ (KB)\\
\midrule
dense\_1000 & no & 1000 & 1000000 & 1000 & 9783 \\
sparsine & yes & 50000 & 799494 & 15 & 10150 \\
crystk03 & yes & 24696 & 887937 & 35 & 10791 \\
ex11 & no & 16614 & 1096948 & 66 & 11004 \\
2cubes\_sphere & yes & 101492 & 874378 & 8 & 11832 \\
xenon1 & no & 48600 & 1181120 & 24 & 12388 \\
raefsky3 & no & 21200 & 1488768 & 70 & 14911 \\
cube2m\_o32 & no & 60044 & 1567463 & 26 & 16774 \\
nasasrb & yes & 54870 & 1366097 & 24 & 16866 \\
cube2m\_n32 & no & 65350 & 1636210 & 25 & 17127 \\
venkat01 & no & 62424 & 1717792 & 27 & 17872 \\
filter3D & yes & 106437 & 1406808 & 13 & 18149 \\
appu & no & 14000 & 1853104 & 132 & 18342 \\
poisson3Db & no & 85623 & 2374949 & 27 & 24697 \\
thermomech\_dK & no & 204316 & 2846228 & 13 & 31386 \\
Ga3As3H12 & yes & 61349 & 3016148 & 49 & 36304 \\
xenon2 & no & 157464 & 3866688 & 24 & 40528 \\
tmt\_sym & yes & 726713 & 2903837 & 3 & 45384 \\
CO & yes & 221119 & 3943588 & 17 & 49668 \\
tmt\_unsym & no & 917825 & 4584801 & 4 & 60907 \\
crankseg\_1 & yes & 52804 & 5333507 & 101 & 63327 \\
SiO2 & yes & 155331 & 5719417 & 36 & 69451 \\
bmw3\_2 & yes & 227362 & 5757996 & 25 & 71029 \\
af\_0\_k101 & yes & 503625 & 9027150 & 17 & 113656 \\
angical & yes & 546587 & 11218066 & 20 & 140002 \\
F1 & yes & 343791 & 13590452 & 39 & 164634 \\
tracer & yes & 1050374 & 14250293 & 13 & 183407 \\
audikw\_1 & yes & 943695 & 39297771 & 41 & 475265 \\
cube2m & no & 2000000 & 52219136 & 26 & 545108 \\
cage15 & no & 5154859 & 99199551 & 19 & 1059358 \\
\bottomrule
\end{tabularx}}
\end{table}

\subsection{Sequential performance}

We have compared the sequential performance of CSRC to the standard CSR.
For symmetric matrices, we have chosen the OSKI implementation \cite{LVDY04a}
as the representative of the symmetric CSR algorithm, assuming that only the
lower part of $A$ is stored.

\begin{figure}[t]
\centering
\includegraphics[height=0.22\textheight]{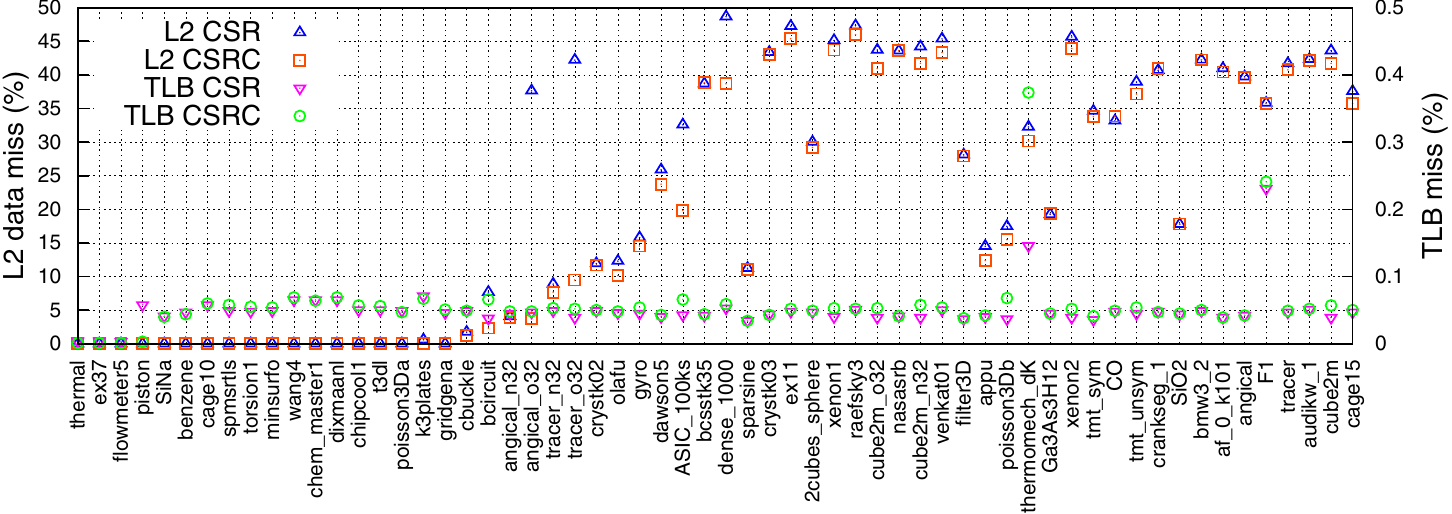}
\caption{Percentages of L2 and TLB cache misses
using CSRC and CSR on the
Wolfdale processor.}
\label{fig:missratio}
\end{figure}

In the sparse matrix-vector product, each element of the matrix is accessed
exactly once.  Thus, accessing these entries incurs only on compulsory misses.
On the other hand, the elements of $x$ and $y$ are accessed multiple times.
This would enable us to take advantage of cache hierachies by
reusing recently accessed values.  In the CSR, the access pattern of the $x$
vector is known to be the major hindrance to the exploitation of data reuse,
because arrays $y$, $ia$, $ja$ and $a$ all have stride-1 accesses. Since the
$y$ vector is not traversed using unit stride anymore in the CSRC, one could argue
that there would be an increase in the number of cache misses.  As
presented in Figure \ref{fig:missratio}, experiments on L2 data cache misses
suggest just the converse, while the ratio of TLB misses is roughly constant.

\begin{figure}[t]
\centering
\includegraphics[height=0.22\textheight]{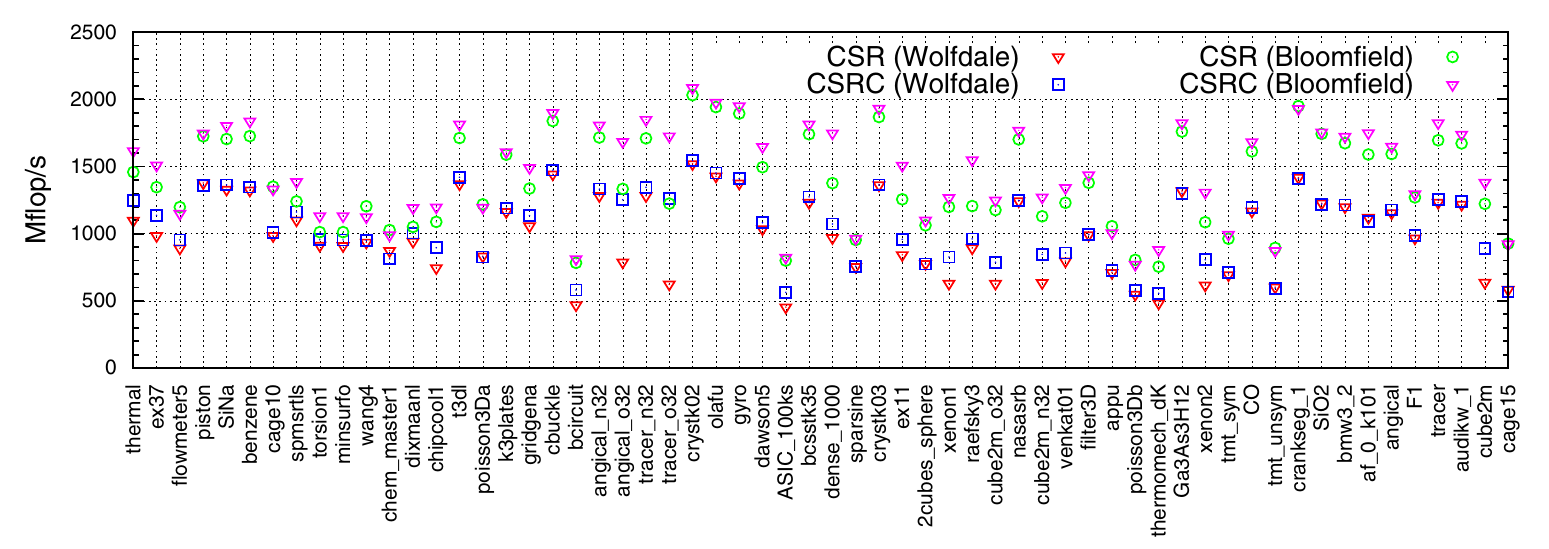}
\caption{Sequential performance in Mflop/s of the matrix-vector
product using CSR and CSRC on both Wolfdale and Bloomfield processors.}
\label{fig:matvec_bench-sequential-mflops}
\end{figure}

The performance of the algorithm considered herein is memory bounded, because the
number of load/store operations is at least as greater as the number of
floating-point multiply-add instructions.  In a dense matrix-vector product, we
need to carry out $O(n^2)$ operations on $O(n^2)$ amount of data, while for sparse
matrices, these quantities are both $O(n)$.
In particular, observe that the computation
of the square $Ax$ product using the CSRC requires the execution of $n$ multiply and $nnz - n$
multiply-add operations, whereas the CSR algorithm requires $nnz$ multiply-add
operations.  On systems without fused multiply-add operations, the CSR and CSRC
algorithms would perform $2nnz$ and $2nnz - n$ floating-point instructions,
respectively.  On the other hand, the number of load instructions for CSR is
$3nnz$, and $\frac{5}{2}nnz - \frac{1}{2}n$ for the CSRC format. Hence the
ratio between loadings and flops is approximately 1.26 for CSRC and exactly 1.5
for CSR.  This bandwidth mitigation may be the most relevant reason for the
efficiency of CSRC shown in Figure \ref{fig:matvec_bench-sequential-mflops}.
It is also worth noting the advantage of the augmented CSRC on
matrices whose square part is numerically symmetric,
i.e., the matrices angical\_o32 and tracer\_o32.

\subsection{Multi-thread version}

Our parallel implementation was evaluated with up to 4 threads on Bloomfield
with Hyper-Threading technology disabled.
The values of speedup
are relative to the pure sequential CSRC algorithm, and not to the one thread case.

\begin{figure}[!t]
\centering
\includegraphics[height=0.22\textheight]{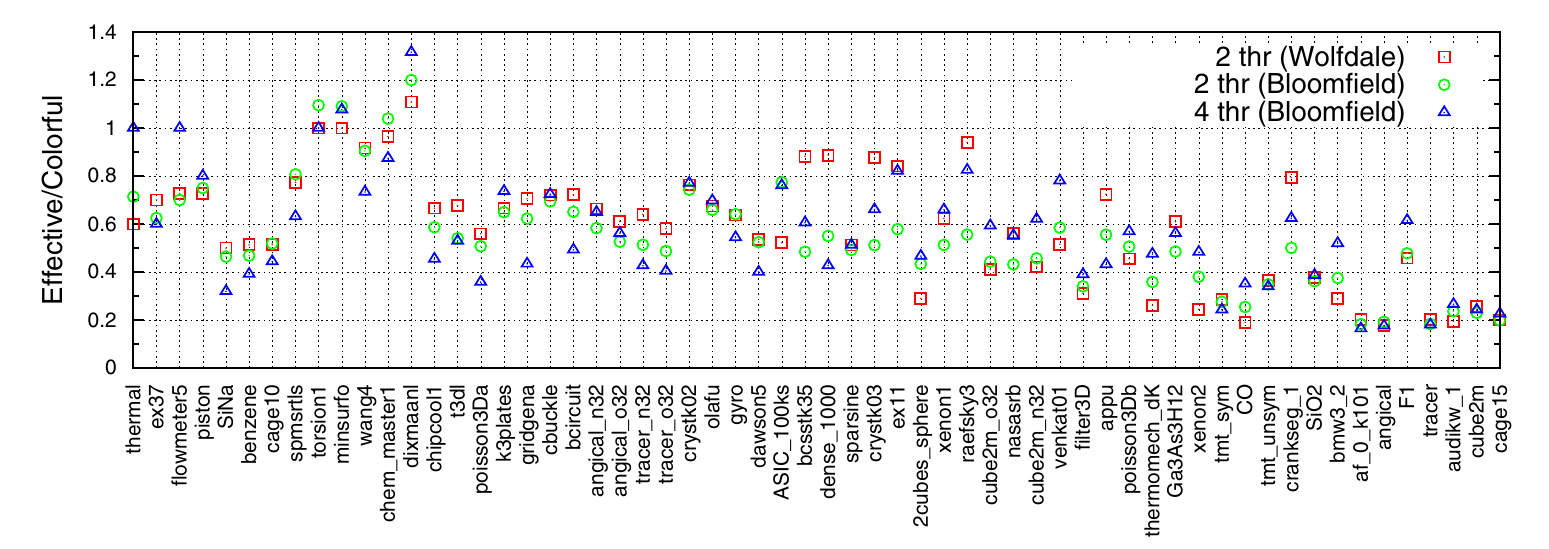}
\caption{Performance comparison between the colorful method and the fastest local buffers implementation
on the Wolfdale and Bloomfield systems.}
\label{fig:matvec_bench-local_buffers_effective_nonzeros_vs_colorful}
\end{figure}

\begin{figure}[!t]
\centering
\begin{tabular}{@{}>{\footnotesize}lm{0.9\textwidth}@{}}
(a) & \subfloat{\includegraphics[height=0.21\textheight]{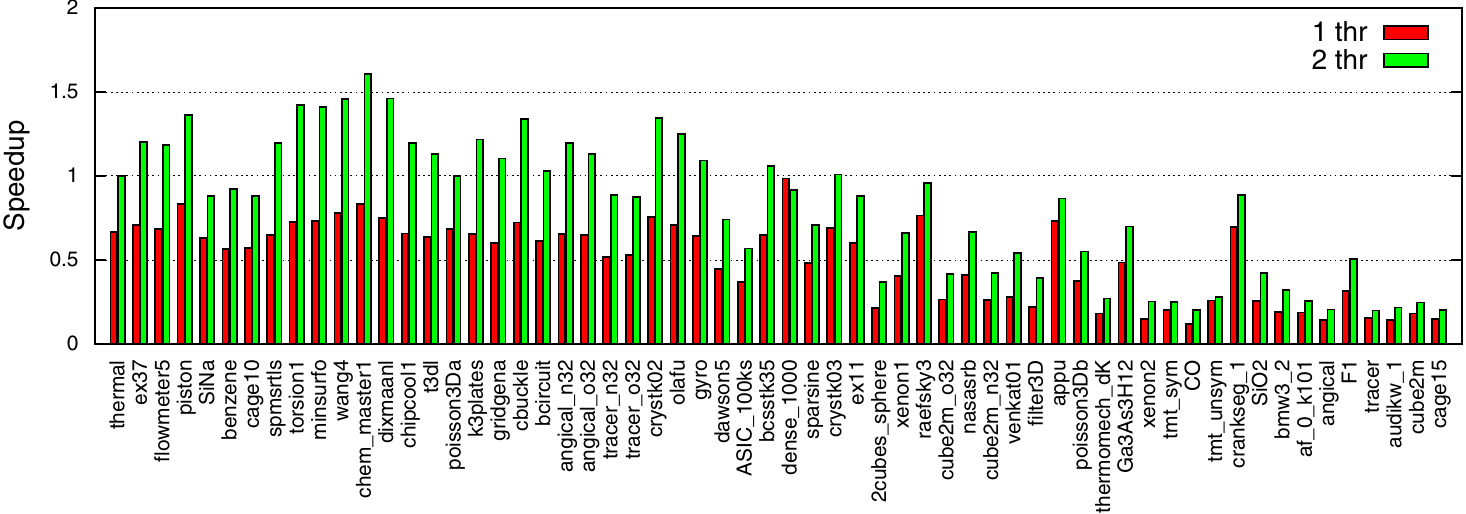}\label{fig:wolfdale-matvec_bench-colorful-speedup}}\\
(b) & \subfloat{\includegraphics[height=0.21\textheight]{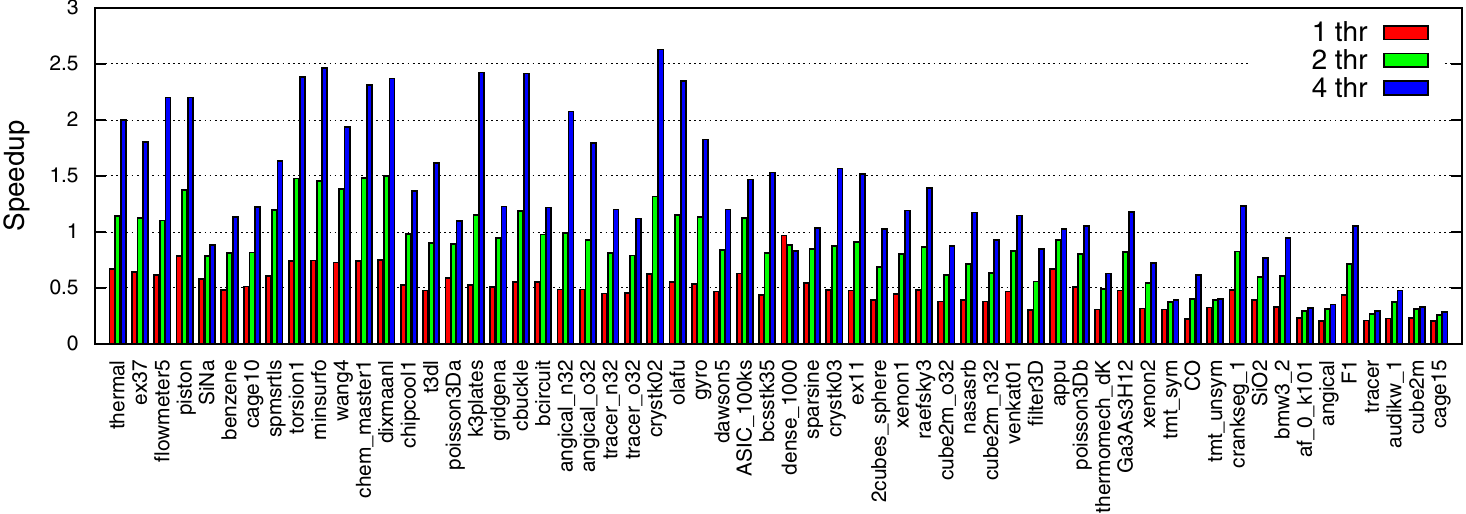}\label{fig:bloomfield-matvec_bench-colorful-speedup}}
\end{tabular}
\caption{Speedups for the colorful method on the (a) Wolfdale and (b) Bloomfield processors.}
\label{fig:matvec_bench-colorful-speedups}
\end{figure}

One would expect that the colorful method is best suited to matrices with
few conflicts, e.g., narrow band matrices,
because the lower is the maximum degree in the conflict graph, the larger is its parallelism.
As shown in Figure
\ref{fig:matvec_bench-local_buffers_effective_nonzeros_vs_colorful},
it was more efficient only on the matrices torsion1, minsurfo and dixmaanl, which have the smallest bandwidth among all matrices.
Nonetheless, according to Figures \subref{fig:wolfdale-matvec_bench-colorful-speedup} and \subref{fig:bloomfield-matvec_bench-colorful-speedup}, small matrices can still benefit from some parallelism.

An important deficiency of the colorful strategy, which contributes to its lack of
locality, is the variable-size stride access to the source and destination
vectors.
Inside each color, there not exist rows
sharing neither $y$ nor $x$ positions, because if they do share there will be a
conflict, therefore they must have different colors.
We claim that there must be an optimal color size to compensate such irregular accesses.

\begin{figure}[p]
\centering
\begin{tabular}{@{}>{\footnotesize}lm{0.9\textwidth}@{}}
(a) & \subfloat{\includegraphics[height=0.21\textheight]{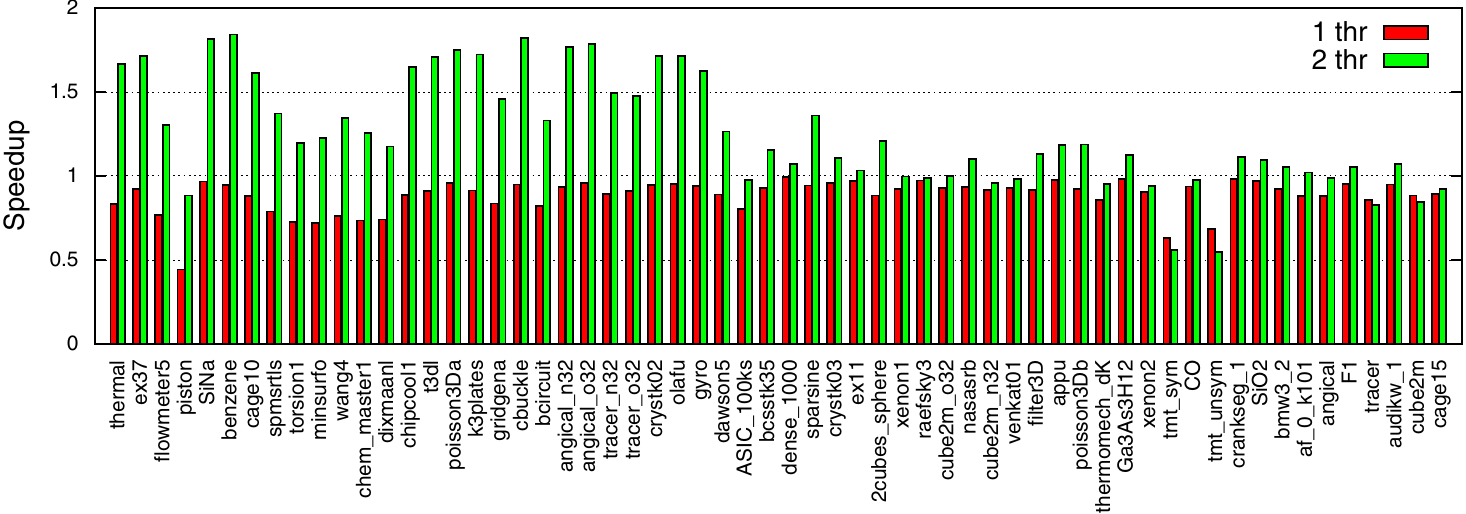}\label{fig:wolfdale-matvec_bench-local_buffers_static_nonzeros-speedup}}\\
(b) & \subfloat{\includegraphics[height=0.21\textheight]{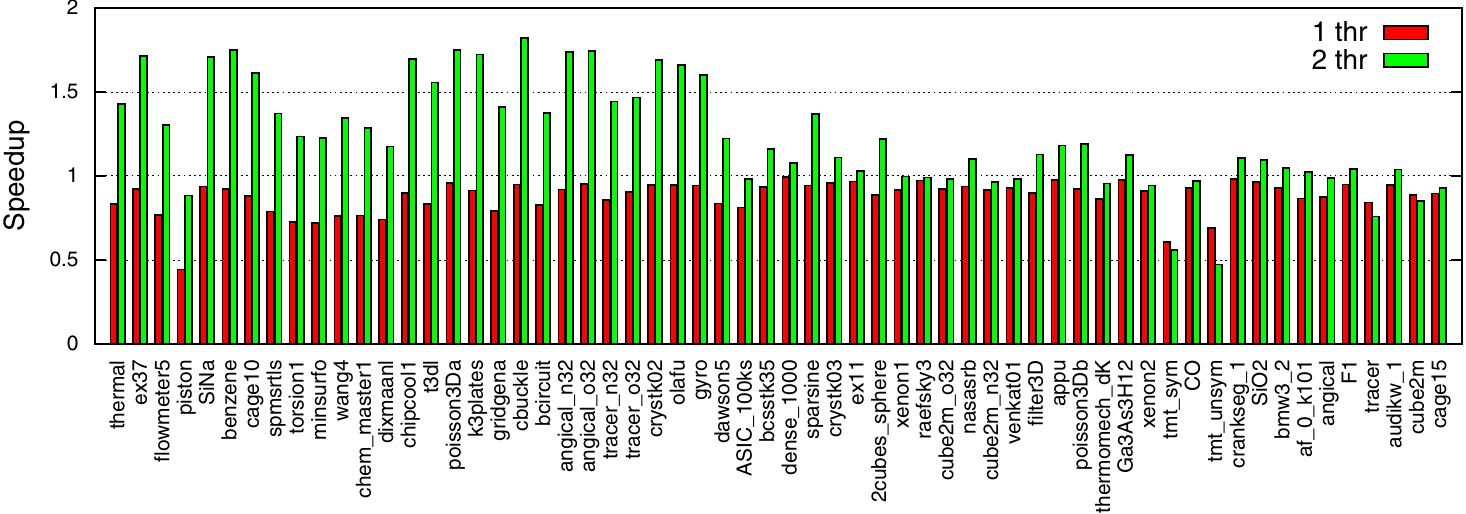}\label{fig:wolfdale-matvec_bench-local_buffers_full_nonzeros-speedup}}\\
(c) & \subfloat{\includegraphics[height=0.21\textheight]{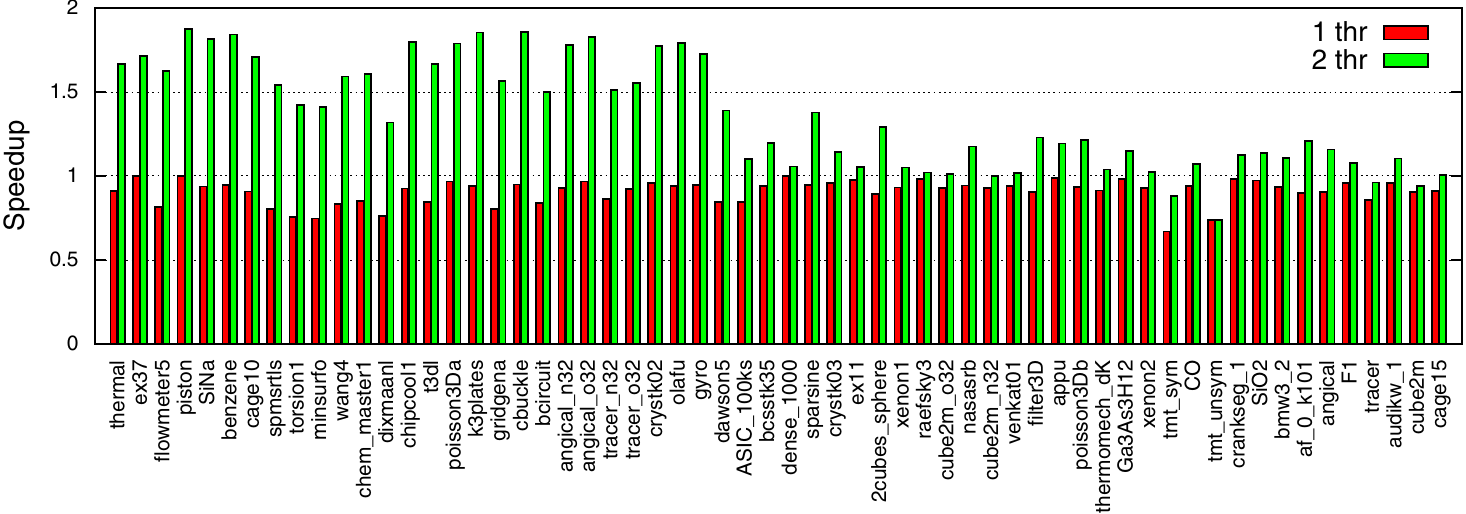}\label{fig:wolfdale-matvec_bench-local_buffers_effective_nonzeros-speedup}}\\
(d) & \subfloat{\includegraphics[height=0.21\textheight]{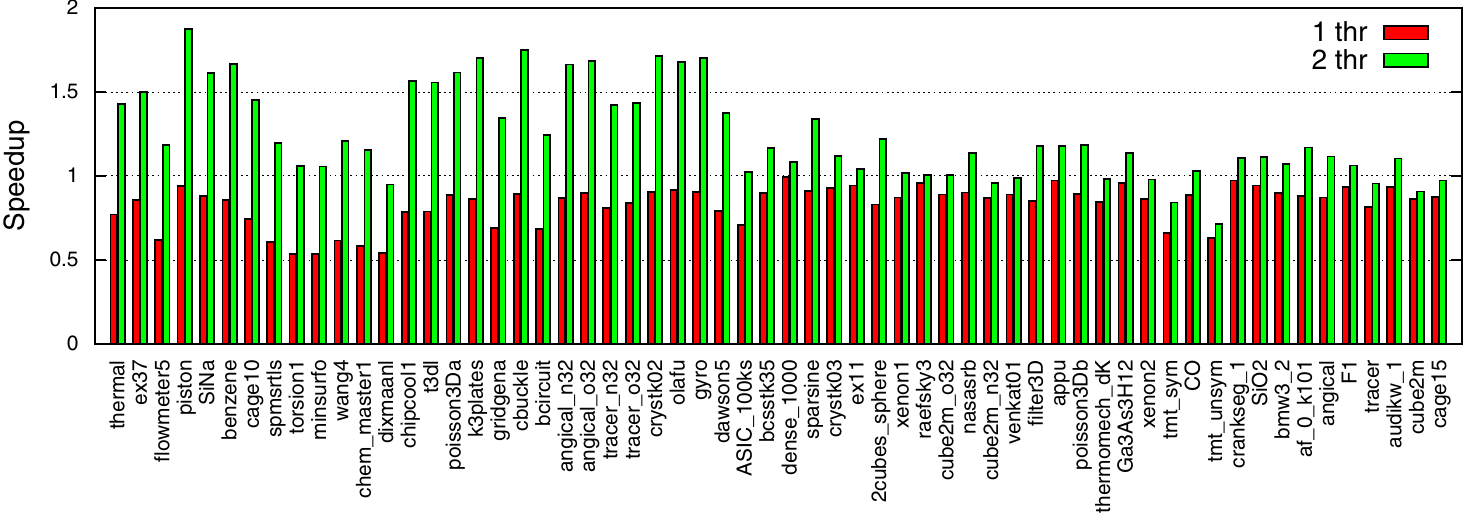}\label{fig:wolfdale-matvec_bench-local_buffers_strict_nonzeros-speedup}}\\
\end{tabular}
\caption{Speedups achieved by the local buffers strategy using the (a) all-in-one, (b) per buffer, (c) effective and (d) interval methods of initialization/accumulation on the Wolfdale processor.}
\label{fig:wolfdale-matvec_bench-local_buffers-speedups}
\end{figure}

\begin{figure}[p]
\centering
\begin{tabular}{@{}>{\footnotesize}lm{0.9\textwidth}@{}}
(a) & \subfloat{\includegraphics[height=0.21\textheight]{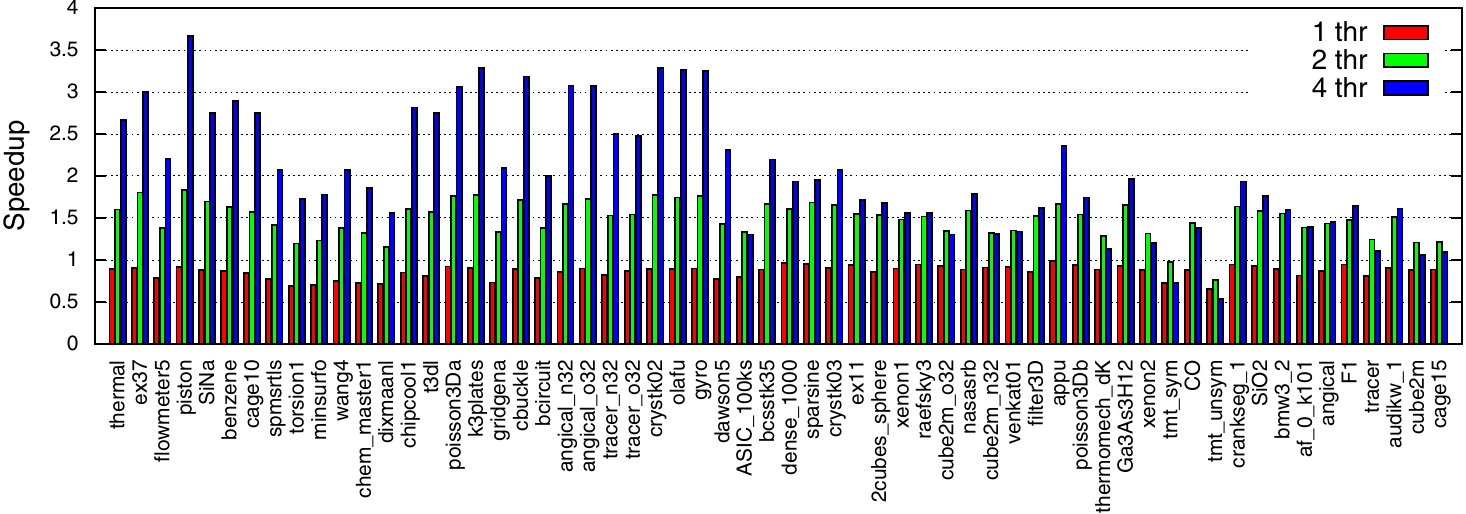}\label{fig:bloomfield-matvec_bench-local_buffers_static_nonzeros-speedup}}\\
(b) & \subfloat{\includegraphics[height=0.21\textheight]{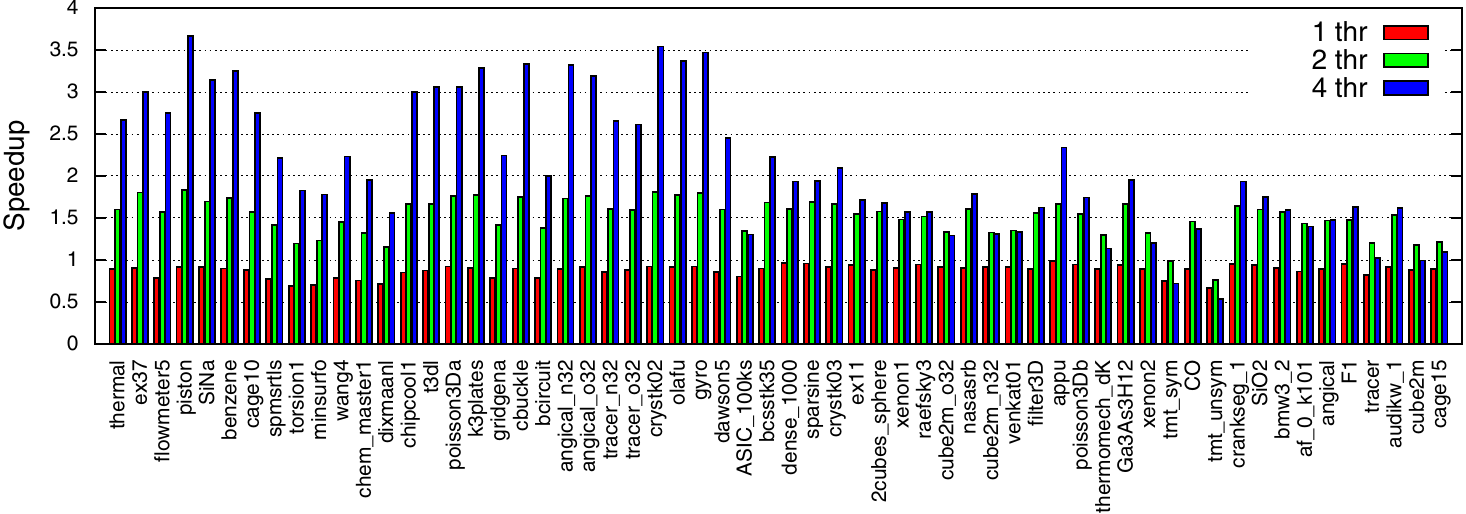}\label{fig:bloomfield-matvec_bench-local_buffers_full_nonzeros-speedup}}\\
(c) & \subfloat{\includegraphics[height=0.21\textheight]{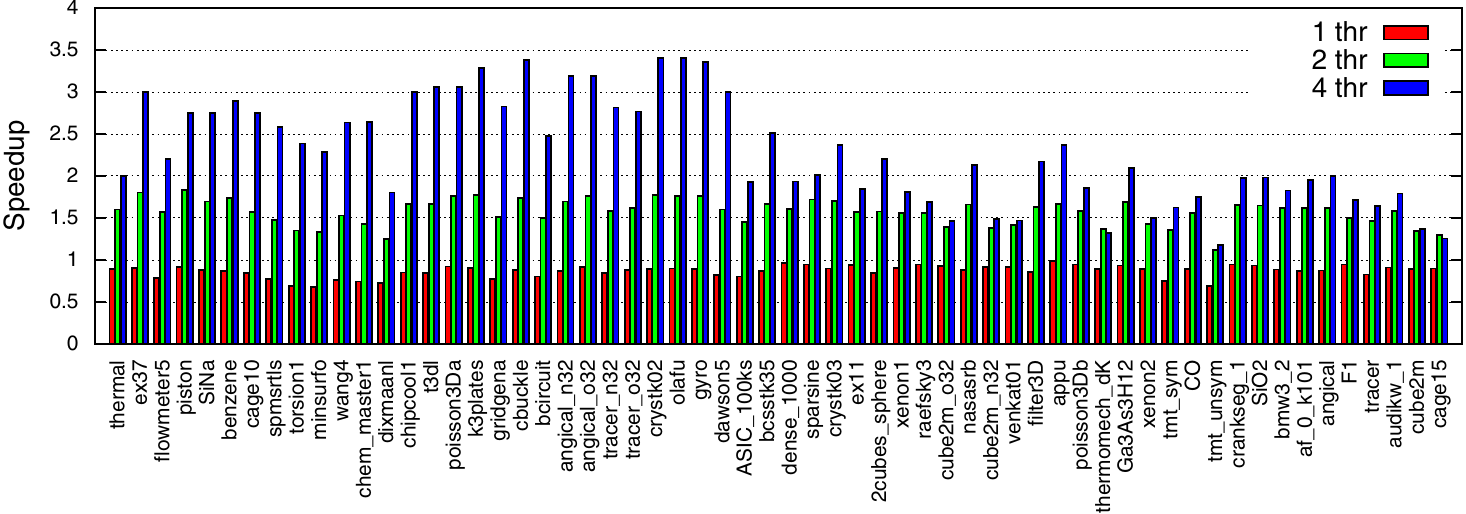}\label{fig:bloomfield-matvec_bench-local_buffers_effective_nonzeros-speedup}}\\
(d) & \subfloat{\includegraphics[height=0.21\textheight]{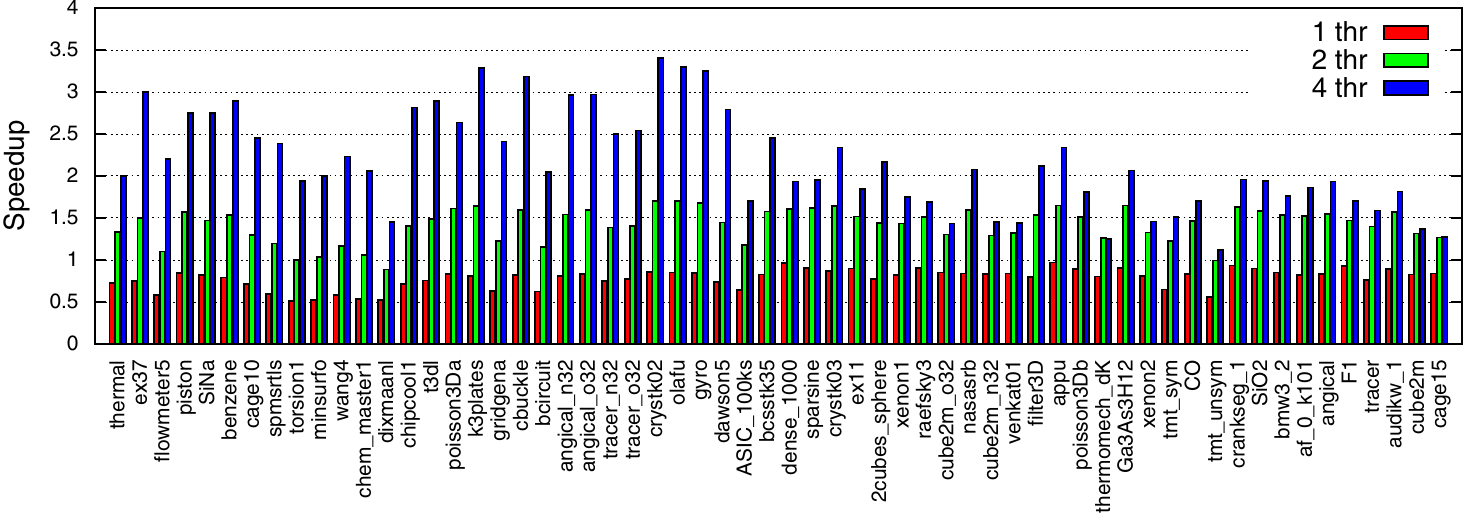}\label{fig:bloomfield-matvec_bench-local_buffers_strict_nonzeros-speedup}}\\
\end{tabular}
\caption{Speedups achieved by the local buffers strategy using the (a) all-in-one, (b) per buffer, (c) effective and (d) interval methods of initialization/accumulation on the Bloomfield processor.}
\label{fig:bloomfield-matvec_bench-local_buffers-speedups}
\end{figure}

Figures \ref{fig:wolfdale-matvec_bench-local_buffers-speedups}
and \ref{fig:bloomfield-matvec_bench-local_buffers-speedups} show the outcomes of speedups attained
by all four implementations of the local buffers strategy.
The
overheads due to the initialization and accumulation steps become
notorious when using just one thread. This can be easily overcome by
checking the number of threads at runtime. If there exists only one thread
in the working team, the global destination vector is used instead.
Although all four implementations reached reasonable speedup peaks, the effective method has been more stable over the whole data set.
On the average, it is the best choice for 93\% of the cases on the Wolfdale, and for 80\% and 78\% on Bloomfield with 2 and 4 threads, respectively.

To better illustrate the performance of different initialization/accumulation algorithms, Table \ref{tab:matvec_bench-acctime} presents average values of the running time consumed by these algorithms considering two classes of matrices, the ones that fit in cache and the others that do not.
As expected, both all-in-one and per buffer strategies have similar performance.
The effective and interval methods have demonstrated to be very feasible for practical use, although the later may incur a higher overheard because the number of intervals is at least as great as the number of threads.

In general, the running time is influenced by the working set size
and the band structure of the matrix.  When the arrays used by the CSRC
fit or nearly fit into cache memory, better speedups were
obtained with almost linear scalability, reaching up to 1.87 on Wolfdale.  For some matrices
from the University of Florida collection it was observed a poor
performance, e.g., tmt\_sym, tmt\_unsym, cage15 and F1.  In the case of cage15 and F1, this may be attributed to the
absence of a band structure.  On
the other hand, there seems to be a bandwidth lower bound for preserving performance.
In particular, the quasi-diagonal profile of the matrices tmt\_sym and tmt\_unsym
have contributed to amplify indirection overheads.

%The main difference between the two testbeds we used is the memory subsystem featured by the Intel Bloomfield processor, namely, the QuickPath technology.
%This may be the key observation for explaining the fact that
Our code has been 63\% more efficient on Bloomfield using 2 threads than on Wolfdale.
Taking a closer view, however, we see that Wolfdale is faster on 80\% of matrices with working set sizes up to 8MB, while Bloomfield beats the former on 94\% of the remaining matrices.
Notice that Wolfdale requires less cycles than Bloomfield to access its outer most cache, which would explain its superiority on small matrices.

Analysing the performance with 4 threads on the Bloomfield processor shown in Figure \subref{fig:bloomfield-matvec_bench-local_buffers_effective_nonzeros-speedup}, speedups indicate that large working sets drastically degrades the efficiency of the implementation, compared to the 2-threaded case.
On smaller matrices, speedups seem to grow linearly, with peaks of 1.83 and 3.40 using 2 and 4 threads, respectively.

\begin{table}[t]
\caption{Average values of the maximum running time among all threads spent during the initialization and accumulation steps using four different approaches.}
\label{tab:matvec_bench-acctime}
\centering
\renewcommand{\arraystretch}{1.1}
{\footnotesize
\begin{tabular}{@{}l@{\ \ }c@{\ \ }cc@{\ \ }cc@{\ \ }c@{}}
\toprule
\multirow{3}{*}{Method} & \multicolumn{2}{c}{Wolfdale} & \multicolumn{4}{c}{Bloomfield}\\\cmidrule(r){2-3}\cmidrule(l){4-7}
 & $ws < 6$MB & $ws > 6$MB & \multicolumn{2}{c}{$ws < 8$MB} & \multicolumn{2}{c}{$ws > 8$MB}\\\cmidrule(r){2-2}\cmidrule(r){3-3}\cmidrule(lr){4-5}\cmidrule(l){6-7}
 & 2 & 2 & 2 & 4 & 2 & 4\\
\midrule
all-in-one & 0.0455 & 4.3831 & 0.0370 & 0.0475 & 1.3127 & 2.5068\\
per buffer & 0.0455 & 4.3876 & 0.0320 & 0.0393 & 1.8522 & 3.8299\\
effective  & \textbf{0.0215} & \textbf{1.8785} & \textbf{0.0176} & \textbf{0.0234} & \textbf{0.8094} & \textbf{1.2575}\\
interval  & 0.0858 & 2.9122 & 0.0748 & 0.0456 & 1.3920 & 1.4939\\
\bottomrule
\end{tabular}}
\end{table}

\section{Conclusion}
\label{sec:conclusion}

We have been concerned with the parallelization of the matrix-vector multiplication
algorithm using the CSRC data structure, focusing on multi-core
architectures.  It has been advocated that multi-core parallelization alone can
compete with purely sequential optimization techniques.  We could observe that,
provided sufficient memory bandwidth, our implementation has demonstrated to be fairly scalable.

The main deficiency of the colorful method is due to variable size stride
accesses, which can destroy any locality provided by matrix reordering
techniques.  We claim that it could be improved by fixing the maximum allowed
stride size inside each color class.  This will be the objective of our future
investigations.

Computing the transpose matrix-vector multiplication is
considered costly when using the standard CSR. An easy but still expensive solution
would be to convert it into the CSC format before spawning threads.
This operation is very straightforward using the CSRC,
as we just need to swap the addresses of $al$ and $au$, and we are done.
Clearly, the computational costs remain the same.

Our results extend previous work
on the computation of the sparse matrix-vector product for
structurally symmetric matrices to multi-core
architectures.
The algorithms hereby presented are now part of a distributed-memory implementation
of the finite element method \cite{RF07a}.
Currently,
we conduct experiments on the effect of coupling both coarse- and fine-grained parallelisms.

\section*{Acknowledgment}

We would like to thank Prof.~Jos{\'e} A.~F.~Santiago and Cid S.~G.~Monteiro for
granting access to the Intel i7 machines used in our experiments. We are also
grateful to Ayd{\i}n Bulu\c{c} and the anonymous reviewers
for the helpful comments.

\bibliographystyle{abbrv}
\bibliography{paper}

\end{document}